# Atomistic insights into the inhomogeneous nature of solute segregation to grain boundaries in magnesium

Risheng Pei[1*], Zhuocheng Xie[1*], Sangbong Yi[2], Sandra Korte-Kerzel[1], Julien Guénolé[3,4], Talal Al-Samman[1]

[1] Institut für Metallkunde und Materialphysik, RWTH Aachen University, D-52056 Aachen, Germany

[2] Institute of Materials and Process Design, Helmholtz-Zentrum Hereon, 21502 Geesthacht, Germany

[3] Université de Lorraine, CNRS, Arts et Métiers, LEM3, 57070 Metz, France

[4] Labex Damas, Université de Lorraine, 57070 Metz, France

## Abstract

In magnesium alloys with multiple substitutional elements, solute segregation at grain boundaries (GBs) has a strong impact on many important material characteristics, such as GB energy and mobility, and therefore, texture. Although it is well established that GB segregation is inhomogeneous, the variation of GB solute composition for random boundaries is still not understood. In the current study, atomic-scale experimental and simulation techniques were used to investigate the compositional inhomogeneity of six different GBs. Three-dimensional atom probe tomography results revealed that GB solute concentration of Nd in Mg varies between 2 to 5 at.%. This variation was not only seen for different GB orientations but also within the GB plane. Correlated atomistic simulations suggest that the inhomogeneous segregation behavior observed experimentally stems from local atomic rearrangements within the GBs and introduce the notion of potential excess free volume in the context of improving the prediction of per-site segregation energies.

The drive towards energy saving and environmental protection is attracting significant interest from research laboratories worldwide to develop innovative and cost-effective materials and processes for lightweight structural components. Owing to their high specific strength and stiffness, magnesium alloys are promising candidates to improve fuel economy and support a sustainable lower carbon vehicle technology [1-3]. Compared to steel or aluminum alloys, processing of magnesium alloys poses challenges related to limited strength, sharp crystallographic textures and plastic anisotropy [4-6], which hinder their widespread commercial usage as rolled products. Overcoming these challenges requires a combination of well-informed processing and micro-alloying strategies in order to depart from basal textures and obtain a favorable alignment of basal planes with the principal deformation direction. This should be combined with grain refinement to increase the material strength, reduce the activity of deformation twinning and promote additional intergranular deformation mechanisms.

Mg alloys containing small additions of rare earth (RE) elements demonstrate remarkable qualitative alterations of the sheet texture along with reduced mechanical anisotropy, and thereby, promoted formability [7-12]. The effectiveness of RE addition for texture softening depends on the type and concentration of the added RE, and can be further increased by co-addition of Zn or Mn. This is known to result in textures with a quadrupole characteristic, i.e. with a distribution of basal poles along the rolling and transverse sheet directions that are favorable for sheet metal forming [7, 13-15]. From a synergistic perspective, multiple solute species with smaller and larger atomic sizes than magnesium tend to co-segregate and form local clusters in order to minimize the lattice misfit with the matrix [13-15]. This raises interest in the characteristics of solute interactions, and the resulting impact on the deformation and recrystallization behavior in terms of active deformation modes and grain boundary migration. Although concrete conclusions regarding the mechanisms of texture modification are still elusive, there is a consensus in the literature that co-segregation of combined solute species inhibits growth of grains with a basal texture by decreasing the grain boundary (GB) energy and mobility [13, 14, 16]. This effect will vary with the type of GB and segregating solute giving rise to a growth preference of certain orientations (e.g. ones with basal pole split in the sheet transverse direction) [7, 14, 16-24].

Given the complexity and experimental limitations to study atomistic behaviors, it is prudent to utilize recent advancements in high performance computing to investigate computationally intensive problems. Atomistic simulations are a powerful complement to high-resolution experimental techniques targeting the atomic-scale behavior of GBs. For example, molecular dynamics (MD) and molecular statics (MS) have been used to compute the distribution of GB segregation energy in face-centered cubic polycrystals [25, 26]. For hexagonal close-packed metals, density functional theory calculations have also been used to study solute segregation at twin boundaries and coincident site lattice (CSL) $\Sigma 7$ GB in magnesium [27, 28]. Another novel application of atomistic simulations is its combination with machine learning to investigate the segregation energetics of aluminum at <0001> symmetric tilt GBs as a function of GB structure and local atomic environment [29].

Despite previous extensive research on the effect of different solutes on recrystallization texture development, formal understanding of inhomogeneous GB segregation and resulting selective growth during recrystallization remains pending. In line with this issue, the present work combines advanced modeling and high-resolution characterization at the atomic-scale to shed light on the effect of GB structure on solute segregation.

The studied material was an extruded Mg-1.0 wt.%Mn-1.0 wt.% Nd alloy (hereafter, MN11) [12, 30]. 3D atom probe tomography (APT) in a Local Electrode Atom Probe 4000X HR from Cameca was employed to quantify the chemical composition of six general GBs using laser-pulsing mode at a temperature of 30 K. Reconstruction of the evaporated tips was performed using the software package IVAS 3.8.2. The APT sample preparation was carried out by transmission Kikuchi diffraction (TKD)-assisted focused ion beam (FIB) milling in a FEI Helios 600i dual-beam electron microscope (Fig. S1). Correlative atomistic simulations to investigate the relationship between per-site segregation energy and the local site environment were performed using the open-source MD software package LAMMPS [31] in conjunction with the modified embedded atom method (MEAM) potential for Mg-Nd [32]. The atomistic configurations of general GBs were constructed using the open source tool Atomsk [33] based on experimentally determined crystallographic orientations of the GB plane and related grains obtained from TKD mapping and reconstructed APT tips (cf. supplementary material).

Fig. 1(a) presents elemental distribution maps of Mg, Mn and Nd atoms from a reconstructed APT tip containing two GB segments with 59.3° $[\bar{3}121]$ and 86.5° $[12\bar{3}1]$ misorientations, and $(70\bar{7}\bar{1}0)$ and $(\bar{2}021)$ boundary planes, respectively. For simplification, the GBs are denoted hereafter by their misorientation angles. Given that Nd has a larger atomic radius (206 pm) than Mg (173 pm), Nd atoms in the solid solution matrix induce compressive elastic strains and therefore tend to segregate at GBs that are rich in microstructural defects. This is depicted in Fig. 1(a) by the obvious Nd enrichment at the two boundaries. The top part of the tip revealed a pure Mn precipitate smaller than 100 nm in diameter, demonstrating evident segregation of Nd atoms at the interface with the matrix. In Fig. 1(b) the obtained concentration profiles along a direction normal to the GB plane (region of interest ROIs 1 & 2) reveal that the segregation level depends on the GB type. The measured Nd peak concentrations at the two boundaries were 2.9 ± 0.2 at.% (86.5° GB) and 3.7 ± 0.2 at.% (59.3° GB).

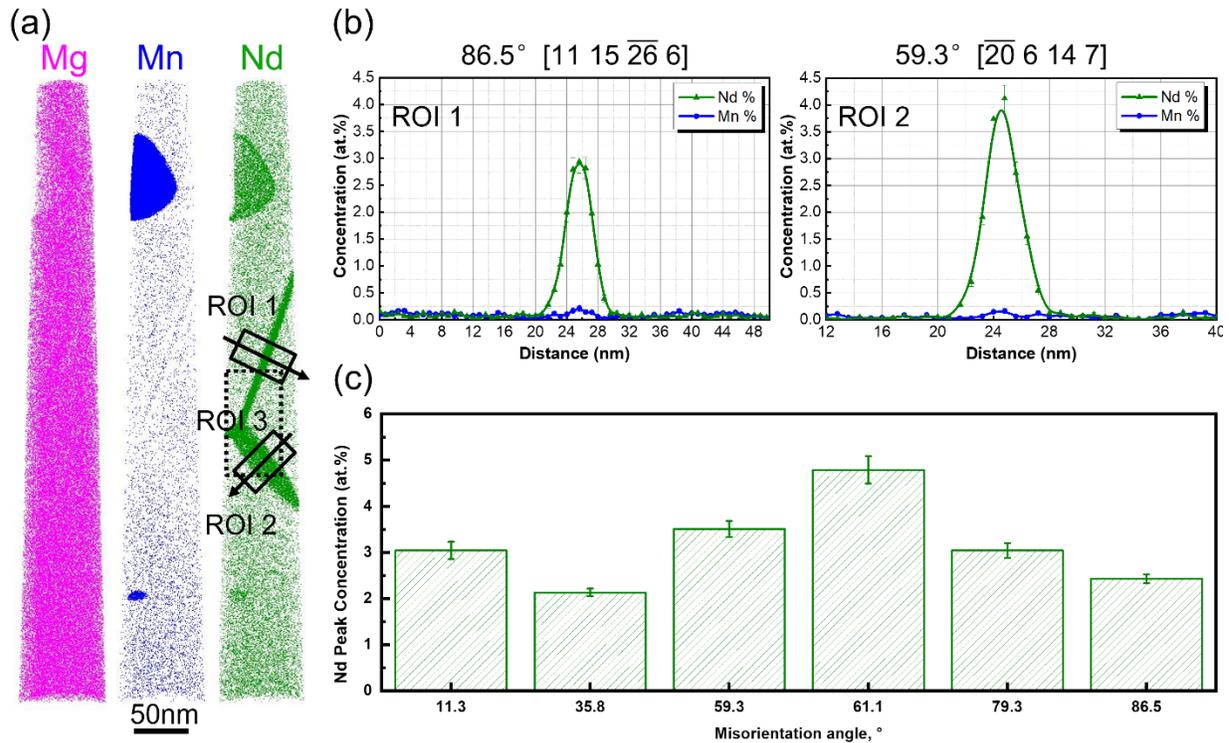

**Fig. 1.** (a) Reconstructed 3D atom distribution maps of Mg, Mn and Nd; (b) solute concentration profiles across the two measured grain boundaries, extracted from cylindrical region of interests ROI 1 and ROI 2 (diameter of 30 nm with a fixed bin size of 0.8 nm) outlined in (a); (c) Variation of the averaged Nd peak concentration in the GB plane for several measured boundaries denoted here by their misorientation angles for simplicity.

A similar trend was also seen in another reconstructed tip containing a different random GB with 11.3° $[10\bar{1}0]$ misorientation and $(\bar{1}\bar{1}47\bar{4})$ boundary plane (Fig. S2). As seen in the atom distribution maps (Fig. S2 (a)), Nd segregation at the GB was more evident than Mn segregation. The corresponding mass spectrum is shown in Fig. S2(b), where Nd peaks can be observed. The concentration profiles of ROI 1 (Fig. S2(c)) indicate peak concentrations of Nd and Mn equal to 3.0 ± 0.2 at.% and 0.3 ± 0.1 at.%, respectively. Compared to the bulk concentration (0.2 ± 0.1 at.%), the GB concentration of Nd was ~15 times higher. To quantify the solubility of Nd atoms at the GB, the in-plane Nd concentrations in the middle of the GB (dashed line in 2D concentration map, Fig. S2(d)) were averaged for each of the six measured GBs. An overview of the averaged peak solute concentrations shown in Fig. 1(c) illustrates the strong variations in the solute concentrations at the GBs with different macroscopic characters.

The segregation behavior of Nd atoms is not only influenced by the macroscopic features of GBs but also the local structural arrangement of atoms in the GB plane. Fig. 2(a) shows a magnified view of the Nd atom map in the triple junction region (ROI 3 in Fig. 1(a)). The pronounced segregation behavior of Nd atoms is displayed using a 1 at.% Nd iso-concentration surface (Fig. 2(b)). As evidenced by the 2D concentration contour map of the same region shown in Fig. 2(c), the local segregation densities of Nd in the x-z plane of the measured tip exhibit strong variation along both GBs. The highest Nd concentration densities were observed at the triple junction and within a distance of 30 nm along 59.3° GB. As in the 59.3° and 86.5° GBs (Fig. 1(c)), the x-z in-plane segregation in the 11.3° GB was similarly inhomogeneous with concentration density variations between 1.5 at.% and 3.5 at.% (Fig. S2(c)). This can also be seen from the 2D Nd concentration density maps of the GB planes (x-y plane) in the three selected GBs, as shown in Fig. 2(d-f).

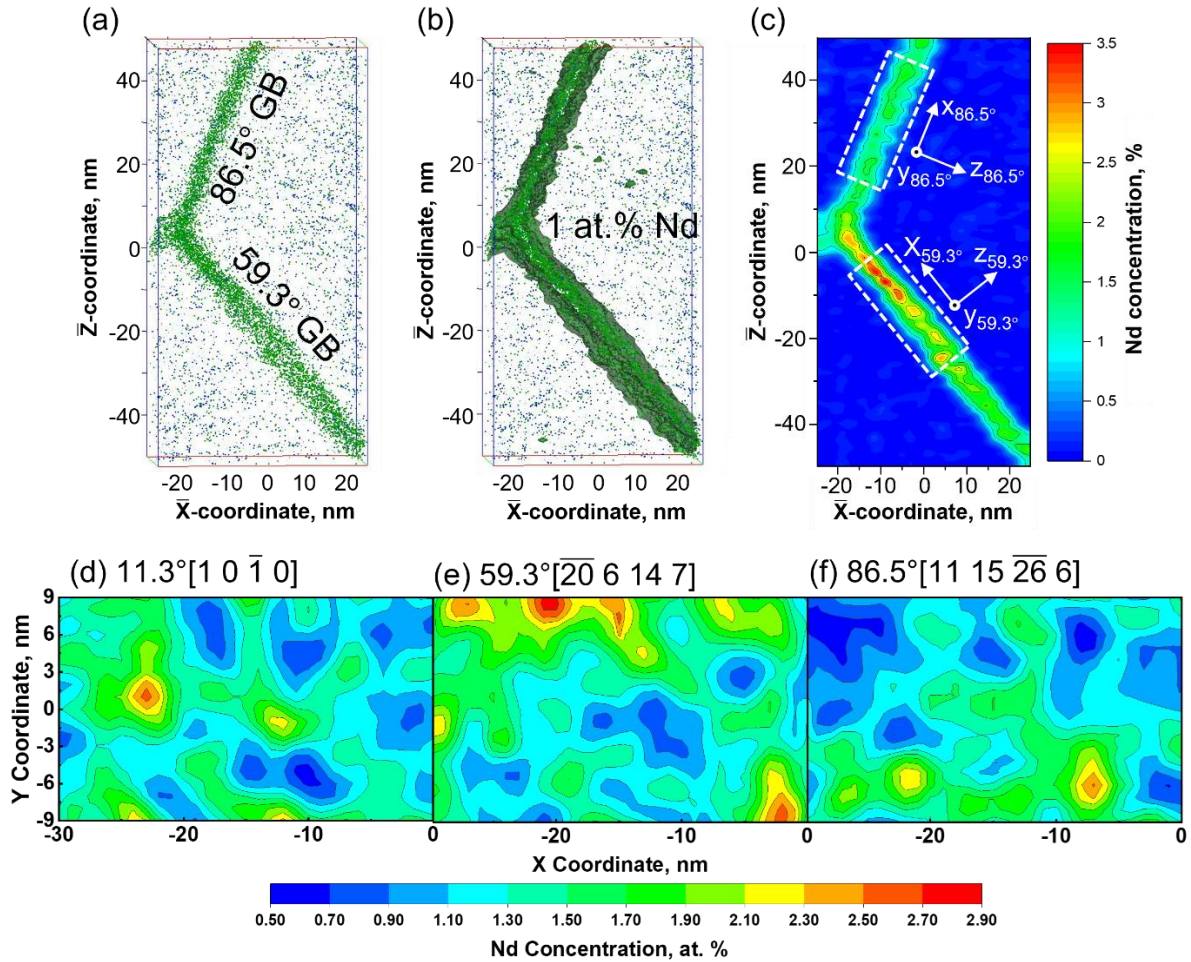

**Fig. 2**. (a) Distribution map of Nd atoms, (b) 1.0 at. % Nd iso-concentration surfaces and (c) 2D-concentration map of Nd in x-z plane of 59.3° and 86.5° GBs in ROI 3 as indicated in Fig. 1(a). Experimental 2D-concentration density maps of in-plane (x-y plane) GB segregation for (d) 11.3° GB in ROI 2 as indicated in Fig. S2(c); (e) 59.3° and (f) 86.5° GBs in the regions highlighted in (c).

To understand the reason for the observed inhomogeneous GB segregation behavior, atomistic simulations on general GBs with experimentally informed characteristics (crystallographic orientations, misorientation angles and GB plane directions) were performed. Due to the low segregation trend of Mn atoms (Figs. 1 and S2), only substitutional Nd segregation at individual GB sites was investigated excluding solute-solute interactions. Two general GBs were selected for the calculations, i.e. 11.3° low angle grain boundary (LAGB) and 59.3° high angle grain boundary (HAGB). A third 86.5° HAGB was also considered, as shown in the supplementary material.

The atomistic configurations were relaxed using the conjugate gradient (with box relaxation in the z-direction) and the FIRE algorithms [34, 35] with force tolerance of $10^{-8}$ eV/Å. A substitution region (80 Å × 80 Å × 20 Å) of Nd substitutions was considered in the center of the cylindrical setup across the GB to neglect the effect of the boundary conditions (Fig. S4). The local site environments within the substitution region represent the possible environments of the GB as indicated in the hydrostatic stress maps in Fig. S5. By swapping one Mg atom with one Nd atom near the GBs in the substitution region, the per-site segregation energies were calculated according to:

$$E_{\text{seg}} = \left(E_{\text{GB}} + E_{\text{bulk}}^{\text{X}}\right) - \left(E_{\text{GB}}^{\text{X}} + E_{\text{bulk}}\right)$$

where $E_{\text{bulk}}$ is the energy of the Mg bulk, $E_{\text{bulk}}^{\text{X}}$ the energy of the Mg bulk where one host atom is replaced by Nd solute, $E_{\text{GB}}$ the energy of the Mg system with a GB, and $E_{\text{GB}}^{\text{X}}$ is the energy of the Mg system with Nd solute occupying a GB site. After each swap, an energy minimization using the FIRE algorithm was performed. The Open Visualization Tool OVITO was used to visualize the atomistic configurations, analyze the misfit dislocation networks and calculate the atomic displacement. The Open Visualization Tool OVITO [36] was used to visualize the atomistic configurations and calculate the atomic displacement. The Dislocation Extraction Algorithm [37] was used to characterize the misfit dislocation networks.

Fig. 3 shows the statistics of per-site segregation energy, which is binned according to the distance to the GB plane. Both GBs exhibit approximately symmetric distributions of segregation energy on either side of the GBs. For most GB sites at a minimum distance of 8 Å from the GB plane center, the segregation energy is close to zero. The 11.3° LAGB in Fig. 3(a) shows distinct segregation behavior compared to the 59.3° HAGB (Fig. 3(b)). The maximum value of the per-site segregation energies of each bin is higher for the 11.3° LAGB than the 59.3° HAGB. For the 11.3° LAGB, the distributions of mean, median and third quartile segregation energies sharply increase when approaching the GB plane, and the deviation between mean and median in each bin is more significant than for the 59.3° HAGB. In contrast, the mean, median and third quartile segregation energies within 2.5 Å of the GB plane of the 59.3° HAGB stay at similar levels (cf. Fig. 3(b)).

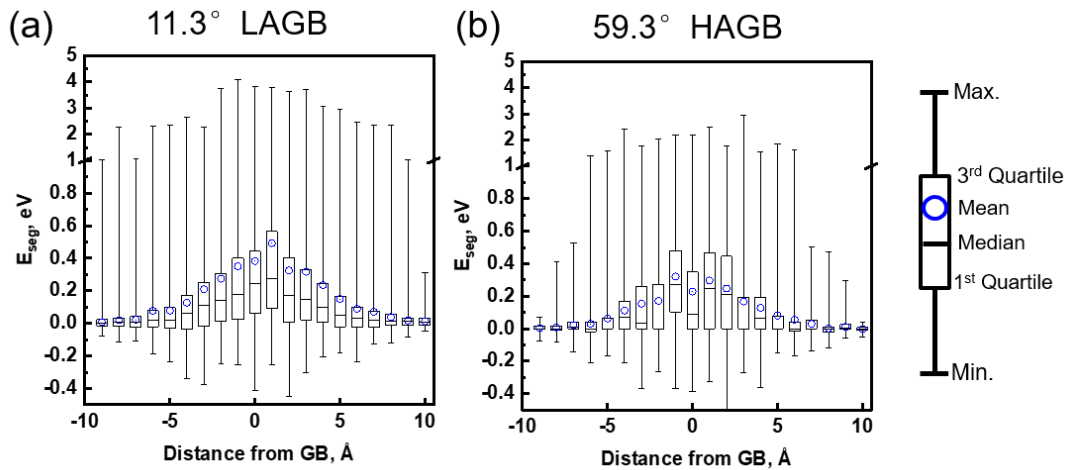

**Fig. 3.** Atomistic simulations of substitutional Nd segregation to (a) 11.3° and (b) 59.3° GBs. Boxplots of segregation energy as a function of distance from the GB center. The data is divided into bins of 1 Å. The upper and lower whisker ends represent maximum and minimum values, respectively. The upper and lower bounds of the boxes are third and first quartiles, respectively. The horizontal lines in the boxes indicate median values. The blue dots represent the mean values.

Fig. 4 displays heat maps of the segregation energy density indicating the in-plane distribution of per-site segregation energies within the two simulated GBs. The 11.3° LAGB in Fig. 4(a) shows more hot spots than the 59.3° HAGB (Fig. 4(b)). In the 11.3° LAGB, the regions with high segregation energy density correlate well with the lines and junctions of misfit dislocation networks (Figs. 4 (a) and S6). Most misfit dislocations are of screw character, since the LAGB is a twist-like GB. This observation agrees with the atomistic simulations from Seki et al. on face-centered cubic [001] twist GBs [38, 39]. Both simulated GBs show a strong inhomogeneous distribution of segregation energy within the GBs plane.

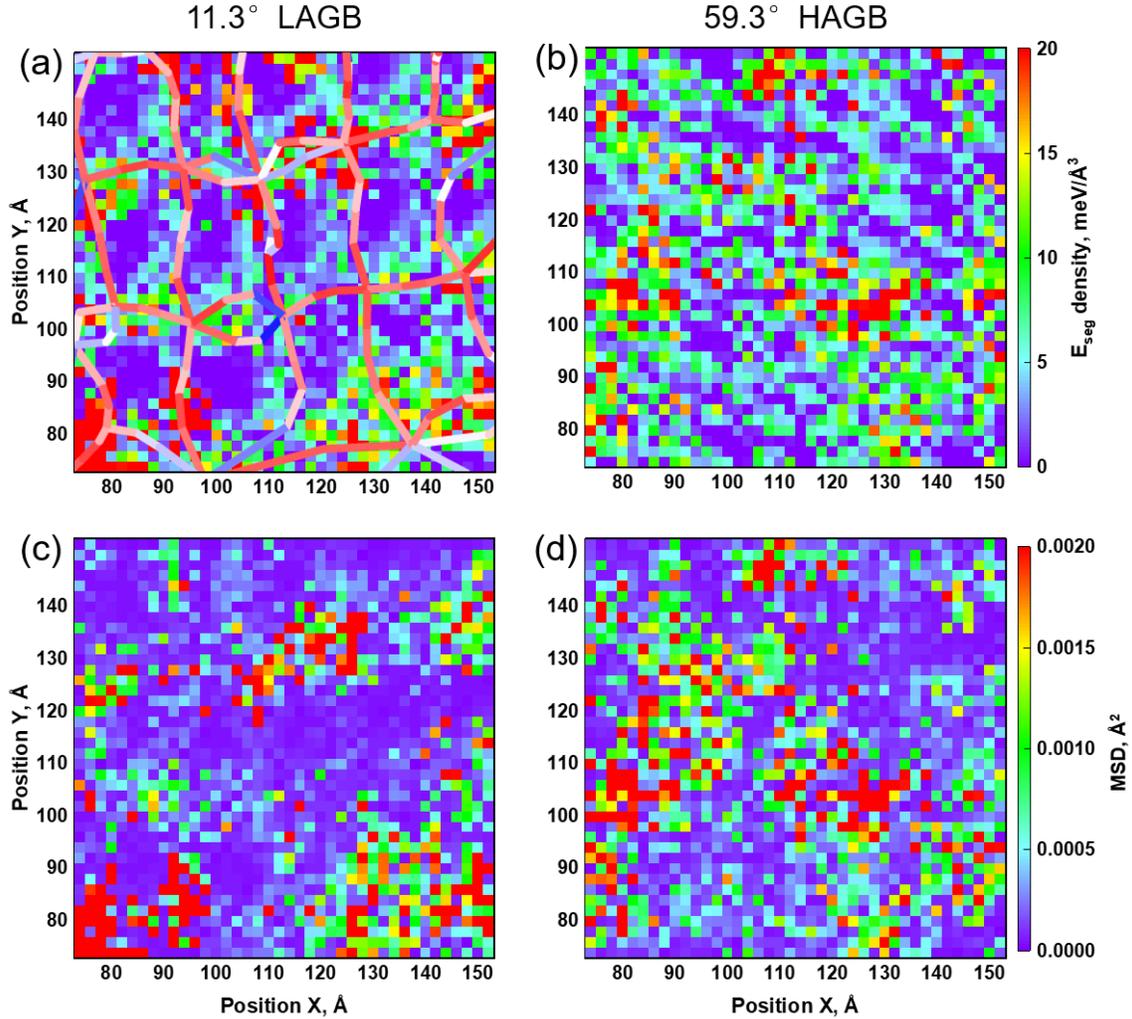

**Fig. 4.** Scatter plots of segregation energy density of the GBs (a, b) and mean squared displacement (MSD) of the GBs with a substitutional Nd atom with respect to the pure Mg counterparts (c, d). The substitution region was divided into 2 Å × 2 Å × 20 Å bins. Only GB sites within a distance of 10 Å from the GB center are calculated. For each bin, mean values of $E_{seg}$ and MSD were obtained. The segregation energy density of each bin was calculated by dividing the mean $E_{seg}$ by the volume of the bin (80 Å³). The GB dislocations in 11.3° LAGB are color coded according to their local character (red: screw; blue: edge).

The concentrations of GB sites with high segregation energy in our simulations as shown in Fig. 4(a, b) agree well with the experimental Nd-solute concentrations at GBs in the current work (Fig. 2(c, d)), and previous experimental observations of solute clusters at HAGBs in Mg-RE solid solutions [24]. The atomistic origin for the inhomogeneous GB segregation observed is explained by correlating the segregation energy to the local structural features of the GB. As shown in Fig. 4(c, d), the distribution of the hot spots of high mean squared displacement (MSD) values correlates well with those of high segregation energy density. For more details of the MSD method, see Supplementary Material. The MSD value at a particular GB site represents the displacements of all the atoms of the sample upon the replacement of the host element by a solute atom at the particular GB site. This indicates the magnitude of local structural rearrangement induced by the introduction of a substitutional solute. A high MSD value suggests that the GB structure can reorganize itself to adapt to the substitutional solute, especially for solute atoms with much larger atomic radius (Nd) than the host element (Mg). This local structural rearrangement renders the site favorable for substitutional segregation. Many studies have shown that the excess free volume is directly related to the GB energy [40] and GB segregation [41, 42], and it was often treated as a macroscopic feature of the GB [43-45]. However, the MSD computed in this work is not related to the excess free volumes of these previous studies. During diffusion, the atoms of the GB reorganize themselves in an energetically favorable configuration that has sufficient free volume to host the solute. The local structural reorganization measured by the MSD acts as a generator of excess free volume. Thus, the MSD is a measure of the potential of excess free volume for a given stable GB configuration, in contrast to the effective excess free volume classically considered. In this work, the strong association between the distributions of per-site segregation energy and MSD demonstrates the impact of the potential excess free volume on per-site segregation energy. The microscopic structural features of a GB, such as GB dislocations, GB disconnections and GB triple junctions, which affect the local site environment, could thus have significant effects on the local segregation behavior. As a result, these features could favor inhomogeneous segregation within the GB, as seen in Fig. 2(a~c) with the high segregation around the triple junction of two boundaries.

Such local structural rearrangement also indicates that the widely used linear elasticity model for the atomistic modelling of high symmetric GBs based on effective excess free volume [19, 20, 28] may not be applicable in the study of general GBs. As shown in Fig. S7, the per-site segregation energy shows a deviation from the predicted segregation energy using the linear elasticity model, especially for sites with high segregation energies. In addition, there is almost no correlation of hot spots between the heat map of the simulated and predicted segregation energy density of the general GBs (see Fig. S6). To improve the prediction of per-site segregation energies in general GBs, existing models should account for the potential excess free volume of GB sites [46, 47].

In summary, the inhomogeneous segregation behavior of Nd solute atoms at several random grain boundaries in a deformed and subsequently annealed MN11 magnesium alloy was investigated by atom probe tomography. Complementary atomistic simulations of per-site segregation energies and mean square displacement of atoms revealed that the inhomogeneous segregation behavior (within the boundary planes and among the different boundaries) originates from the local atomic arrangement within the structural units of grain boundaries. Our results can contribute to improving the understanding of the role of inhomogeneous segregation on GB mobilities in magnesium alloys.

## Acknowledgements

R.P., S.Y. and T.A.S. are grateful for the financial support from the German Research Foundation (DFG) (Grant Nr. AL1343/7-1, AL1343/8-1 and Yi 103/3-1). Z.X. and T.A.S. acknowledge the financial support by the DFG (Grant Nr. 505716422). Z.X. and S.K.K. acknowledge financial support by the DFG through the projects A02, A05 and C02 of the SFB1394 Structural and Chemical Atomic Complexity – From Defect Phase Diagrams to Material Properties, project ID 409476157. Additionally, Z.X. and S.K.K. are grateful for funding from the European Research Council (ERC) under the European Union's Horizon 2020 research and innovation programme (grant agreement No. 852096 FunBlocks). J.G. acknowledges funding from the French National Research Agency (ANR), Grant ANR-21-CE08-0001 (ATOUUM) and ANR-22-CE92-0058-01 (SILA). Simulations were performed with computing resources granted by RWTH Aachen University under project (rwth0591) and by the EXPLOR center of the Université de Lorraine and by the GENCI-TGCC (Grant 2020-A0080911390). We want to thank Dr. Liam Huber and Dr. Dimitri Chauraud (MPIE Düsseldorf) for fruitful discussions and Hexin Wang (RWTH Aachen) for his help with potential benchmark.

## Reference

[1] M.R. Barnett, M.D. Nave, C.J. Bettles, Deformation microstructures and textures of some cold rolled Mg alloys, Mater. Sci. Eng., A 386(1-2) (2004) 205-211.
[2] I.J. Polmear, Magnesium alloys and applications, Mater. Sci. Technol. 10(1) (1994) 1-16.
[3] M.T. Pérez-Prado, O.A. Ruano, Texture evolution during annealing of magnesium AZ31 alloy, Scripta Mater. 46(2) (2002) 149-155.
[4] Y. Chino, M. Kado, M. Mabuchi, Compressive deformation behavior at room temperature-773 K in Mg-0.2 mass%(0.035at.%)Ce alloy, Acta Mater. 56(3) (2008) 387-394.
[5] Y. Chino, M. Kado, M. Mabuchi, Enhancement of tensile ductility and stretch formability of magnesium by addition of 0.2 wt%(0.035 at%)Ce, Mater. Sci. Eng., A 494(1-2) (2008) 343-349.
[6] M.H. Yoo, Slip, twinning, and fracture in hexagnal closed-packed metals, Metall. Trans. A 12(3) (1981) 409-418.
[7] I. Basu, T. Al-Samman, Triggering rare earth texture modification in magnesium alloys by addition of zinc and zirconium, Acta Mater. 67 (2014) 116-133.

[8] T. Al-Samman, X. Li, Sheet texture modification in magnesium-based alloys by selective rare earth alloying, Mater. Sci. Eng., A 528(10-11) (2011) 3809-3822.
[9] K. Hantzsche, J. Bohlen, J. Wendt, K.U. Kainer, S.B. Yi, D. Letzig, Effect of rare earth additions on microstructure and texture development of magnesium alloy sheets, Scripta Mater. 63(7) (2010) 725-730.
[10] X. Li, T. Al-Samman, S. Mu, G. Gottstein, Texture and microstructure development during hot deformation of ME20 magnesium alloy: Experiments and simulations, Mater. Sci. Eng., A 528(27) (2011) 7915-7925.
[11] L.W.F. Mackenzie, M.O. Pekguleryuz, The recrystallization and texture of magnesium-zinc-cerium alloys, Scripta Mater. 59(6) (2008) 665-668.
[12] S.K. Woo, R. Pei, T. Al-Samman, D. Letzig, S. Yi, Plastic instability and texture modification in extruded Mg-Mn-Nd alloy, J. Magnes. Alloy. (2021).
[13] R. Pei, Y. Zou, D. Wei, T. Al-Samman, Grain boundary co-segregation in magnesium alloys with multiple substitutional elements, Acta Mater. 208 (2021) 116749.
[14] R. Pei, Y. Zou, M. Zubair, D. Wei, T. Al-Samman, Synergistic effect of Y and Ca addition on the texture modification in AZ31B magnesium alloy, Acta Mater. 233 (2022) 117990.
[15] F. Mouhib, R. Pei, B. Erol, F. Sheng, S. Korte-Kerzel, T. Al-Samman, Synergistic effects of solutes on active deformation modes, grain boundary segregation and texture evolution in Mg-Gd-Zn alloys, Mater. Sci. Eng., A 847 (2022) 143348.
[16] Z.R. Zeng, Y.M. Zhu, S.W. Xu, M.Z. Bian, C.H.J. Davies, N. Birbilis, J.F. Nie, Texture evolution during static recrystallization of cold-rolled magnesium alloys, Acta Mater. 105 (2016) 479-494.
[17] I. Basu, T. Al-Samman, G. Gottstein, Shear band-related recrystallization and grain growth in two rolled magnesium-rare earth alloys, Mater. Sci. Eng., A 579 (2013) 50-56.
[18] I. Basu, K.G. Pradeep, C. Mießen, L.A. Barrales-Mora, T. Al-Samman, The role of atomic scale segregation in designing highly ductile magnesium alloys, Acta Mater. 116 (2016) 77-94.
[19] H. Xie, Q. Huang, J. Bai, S. Li, Y. Liu, J. Feng, Y. Yang, H. Pan, H. Li, Y. Ren, G. Qin, Nonsymmetrical Segregation of Solutes in Periodic Misfit Dislocations Separated Tilt Grain Boundaries, Nano Lett. 21(7) (2021) 2870-2875.
[20] H. Xie, H. Pan, J. Bai, D. Xie, P. Yang, S. Li, J. Jin, Q. Huang, Y. Ren, G. Qin, Twin Boundary Superstructures Assembled by Periodic Segregation of Solute Atoms, Nano Lett. 21(22) (2021) 9642-9650.
[21] M. Bugnet, A. Kula, M. Niewczas, G.A. Botton, Segregation and clustering of solutes at grain boundaries in Mg-rare earth solid solutions, Acta Mater. 79 (2014) 66-73.
[22] J.F. Nie, Y.M. Zhu, J.Z. Liu, X.Y. Fang, Periodic Segregation of Solute Atoms in Fully Coherent Twin Boundaries, Science 340(6135) (2013) 957-960.
[23] Y.M. Zhu, M.Z. Bian, J.F. Nie, Tilt boundaries and associated solute segregation in a Mg–Gd alloy, Acta Mater. 127 (2017) 505-518.
[24] M. Bugnet, A. Kula, M. Niewczas, G.A. Botton, Segregation and clustering of solutes at grain boundaries in Mg–rare earth solid solutions, Acta Mater. 79 (2014) 66-73.
[25] M. Wagih, C.A. Schuh, Spectrum of grain boundary segregation energies in a polycrystal, Acta Mater. 181 (2019) 228-237.
[26] M. Wagih, C.A. Schuh, Grain boundary segregation beyond the dilute limit: Separating the two contributions of site spectrality and solute interactions, Acta Mater. 199 (2020) 63-72.
[27] Z. Pei, R. Li, J.-F. Nie, J.R. Morris, First-principles study of the solute segregation in twin boundaries in Mg and possible descriptors for mechanical properties, Materials & Design 165 (2019) 107574.
[28] L. Huber, J. Rottler, M. Militzer, Atomistic simulations of the interaction of alloying elements with grain boundaries in Mg, Acta Mater. 80 (2014) 194-204.
[29] J. Messina, R. Luo, K. Xu, G. Lu, H. Deng, M.A. Tschopp, F. Gao, Machine learning to predict aluminum segregation to magnesium grain boundaries, Scripta Mater. 204 (2021) 114150.
[30] R. Pei, S.K. Woo, S. Yi, T. Al-Samman, Effect of solute clusters on plastic instability in magnesium alloys, Mater. Sci. Eng., A 835 (2022) 142685.
[31] A.P. Thompson, H.M. Aktulga, R. Berger, D.S. Bolintineanu, W.M. Brown, P.S. Crozier, P.J. in 't Veld, A. Kohlmeyer, S.G. Moore, T.D. Nguyen, R. Shan, M.J. Stevens, J. Tranchida, C. Trott, S.J. Plimpton, LAMMPS - a flexible simulation tool for particle-based materials modeling at the atomic, meso, and continuum scales, Comput. Phys. Commun. 271 (2022) 108171.
[32] K.-H. Kim, B.-J. Lee, Modified embedded-atom method interatomic potentials for Mg-Nd and Mg-Pb binary systems, Calphad 57 (2017) 55-61.

[33] P. Hirel, Atomsk: A tool for manipulating and converting atomic data files, Comput. Phys. Commun. 197 (2015) 212-219.
[34] E. Bitzek, P. Koskinen, F. Gähler, M. Moseler, P. Gumbsch, Structural Relaxation Made Simple, Phys. Rev. Lett. 97(17) (2006) 170201.
[35] J. Guénolé, W.G. Nöhring, A. Vaid, F. Houllé, Z. Xie, A. Prakash, E. Bitzek, Assessment and optimization of the fast inertial relaxation engine (fire) for energy minimization in atomistic simulations and its implementation in lammps, Computational Materials Science 175 (2020) 109584.
[36] A. Stukowski, Visualization and analysis of atomistic simulation data with OVITO–the Open Visualization Tool, Modell. Simul. Mater. Sci. Eng. 18(1) (2009) 015012.
[37] A. Stukowski, V.V. Bulatov, A. Arsenlis, Automated identification and indexing of dislocations in crystal interfaces, Modell. Simul. Mater. Sci. Eng. 20(8) (2012) 085007-1-085007-16.
[38] A. Seki, D.N. Seidman, Y. Oh, S.M. Foiles, Monte Carlo simulations of segregation at [001] twist boundaries in a Pt(Au) alloy—I. Results, Acta Metall. Mater. 39(12) (1991) 3167-3177.
[39] A. Seki, D.N. Seidman, Y. Oh, S.M. Foiles, Monte Carlo simulations of segregation at [001] twist boundaries in a Pt(Au) alloy—II. Discussion, Acta Metall. Mater. 39(12) (1991) 3179-3185.
[40] A. Seeger, G. Schottky, Die energie und der elektrische widerstand von grosswinkelkorngrenzen in metallen, Acta Metall. 7(7) (1959) 495-503.
[41] Z. Huang, F. Chen, Q. Shen, L. Zhang, T.J. Rupert, Combined effects of nonmetallic impurities and planned metallic dopants on grain boundary energy and strength, Acta Mater. 166 (2019) 113-125.
[42] H. Mehrer, Diffusion in Solids, Springer, Berlin, Heidelberg2007.
[43] E.-M. Steyskal, B. Oberdorfer, W. Sprengel, M. Zehetbauer, R. Pippan, R. Würschum, Direct Experimental Determination of Grain Boundary Excess Volume in Metals, Phys. Rev. Lett. 108(5) (2012) 055504.
[44] J.J. Bean, K.P. McKenna, Origin of differences in the excess volume of copper and nickel grain boundaries, Acta Mater. 110 (2016) 246-257.
[45] H. Sun, C.V. Singh, Temperature dependence of grain boundary excess free volume, Scripta Mater. 178 (2020) 71-76.
[46] L. Huber, B. Grabowski, M. Militzer, J. Neugebauer, J. Rottler, Ab initio modelling of solute segregation energies to a general grain boundary, Acta Mater. 132 (2017) 138-148.
[47] R. Mahjoub, N. Stanford, The electronic origins of the "rare earth" texture effect in magnesium alloys, Scientific Reports 11(1) (2021) 14159.

# Supplementary Material:

# Atomistic insights into the inhomogeneous nature of solute segregation to grain boundaries in magnesium

Risheng Pei[1*], Zhuocheng Xie[1*], Sangbong Yi[2], Sandra Korte-Kerzel[1], Julien Guénolé[3,4], Talal Al-Samman[1]

[1] Institut für Metallkunde und Materialphysik, RWTH Aachen University, D-52056 Aachen, Germany

[2] Institute of Materials and Process Design, Helmholtz-Zentrum Hereon, 21502 Geesthacht, Germany

[3] Université de Lorraine, CNRS, Arts et Métiers, LEM3, 57070 Metz, France

[4] Labex Damas, Université de Lorraine, 57070 Metz, France

## Methods

### TKD-EBSD assisted preparation of APT tips

Before the milling process, the orientation of grains of the sampled surface is characterized by electron backscatter diffraction (EBSD) performed in a FEI Helios 600i dual-beam scanning electron microscope/focused ion beam (SEM/FIB) with an operation voltage of 20 kV, as shown in Fig. S1 a. The specimens for EBSD measurements were prepared by conventional mechanical grinding and polishing followed by electro-polishing in Struers AC-2 reagent at 20 V for 90 s. The targeted grain boundaries were selected based on the EBSD orientation measurements and marked by Pt deposition before they were lifted-out (Fig. S1 b).

To guide the site-specific preparation process and ensure a proper position of the targeted grain boundary within the APT tips, transmission Kikuchi diffraction EBSD (TKD-EBSD) in conjunction with FIB milling at 30 kV with a current of 5.5 nA was employed as shown in Fig. S1 c for thinning steps between 750 and 300 nm inner diameters. After the final low-energy milling step at 2 kV, the targeted GB was ~ 200 nm away from the top of the tip.

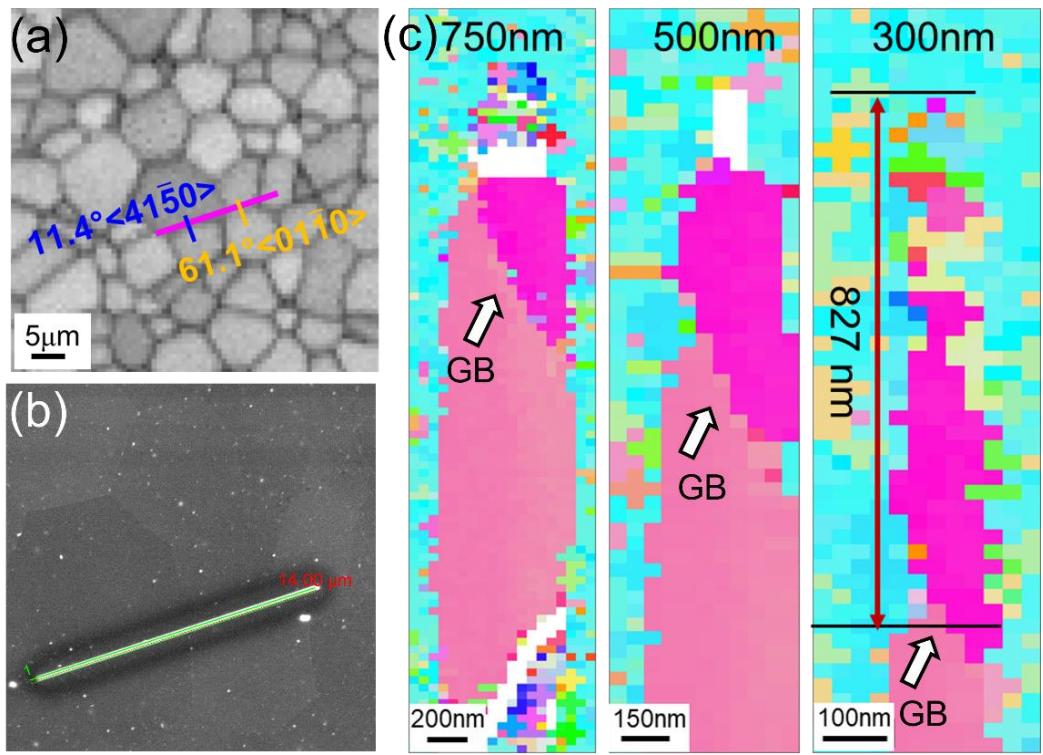

**Fig. S1** Illustration of the TKD-EBSD guided FIB milling for site-specific APT sample preparation: (a) EBSD confidence index (CI) map to select the targeted GBs; (b) secondary electron (SE) image of a targeted grain boundaries from (a) marked by Pt deposition; (c) TKD-EBSD maps during the milling process taken at 30 kV illustrating the misorientation information of grain boundary and the depth of the boundary from top of the tip.

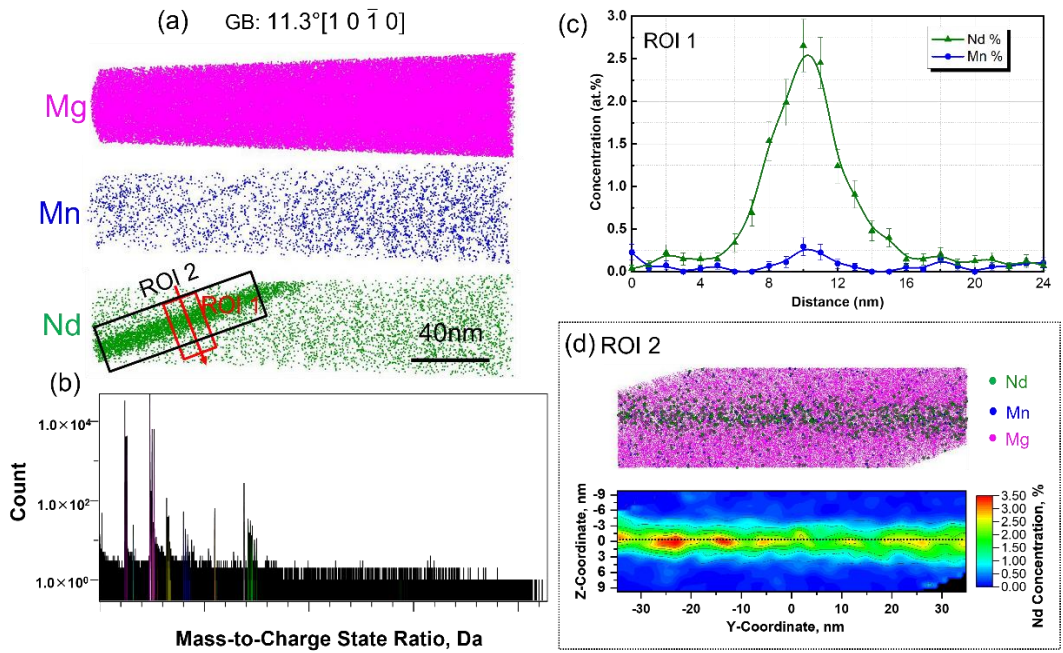

**Fig. S2** Reconstructed 3D APT tip of the MN11 alloy with a long segment of a detected GB milled by guidance of TKD shown in Fig. S: (a) elemental distributions of Mg, Mn and Nd; (b) atom probe mass spectrum, showing distinct peaks of Nd atoms; (c) concentration profile across the GB, extracted from a cylindrical ROI 1(diameter of 15 nm with a fixed bin size of 0.8 nm) outlined in (a); (d) atom distribution (Mg, Mn and Nd) and 2D Nd concentration along X-axis shown in the atom extracted from a cubic ROI 2 outlined in (a).

**Atomistic simulations**

The atomistic simulations were performed using the open-source MD software package LAMMPS [1]. The interatomic interactions were modeled by the modified embedded atom method (MEAM) potential for Mg-Nd by Kim et al. [2]. The material properties of the MEAM potential were benchmarked, particularly the per-site segregation energies at the selected grain boundaries. The results are in good agreement with the experimental and ab-initio data (as presented in Table S1) [3-6]. In addition, we calculated the per-site segregation energy of Nd solute at Σ7 21.8° {12$\bar{3}$0}<0001> grain boundary, {10$\bar{1}$1} and {10$\bar{1}$2} twin boundaries and compared with the ab-initio calculations [6, 7]. The results shown in Fig. S3 prove the potential can describe the segregation trend of Nd solute at the Mg grain boundaries in reasonable agreement with the ab-initio results. The atomistic configurations of general GBs were constructed using Atomsk [8] based on the experimentally-informed crystallographic orientations of two grains and the GB plane (normal to the z-axis) obtained from TKD grain mapping and reconstructed APT tips. The detailed information the boundaries are listed in Table S2. Two crystals were merged along the z-axis and positioned in a cylindrical configuration with a diameter of 22.4 nm equal to the height. The cylinder axis was aligned with the z-axis of the crystals. The GB sitting in the middle of the sample was then lifted out. One atom from each pair of atoms within a distance of separation of 2 Å (62.5% of first nearest neighbor distance) was considered as an overlapping atom and was deleted. Since there is no periodic repeat distance for general GBs, the in-plane scanning over all possible translations of one crystal relative to the other, which is routinely applied in the study of CSL GBs, is intractable. An alternative way to optimize the general GB structures is to vary the deletion distance of overlapping atoms near the interface [9]. Since we focus on per-site segregation energies and local GB structures instead of global GB properties, only one deletion distance was chosen in this work. A reasonable distribution of local site environments, which samples the space of possible environments similar to the minimum energy GB is expected [9]. A schematic of the cylindrical setup is shown in Fig. S4. The top and bottom layers of the cylinder with a thickness of 1.2 nm (2 times interatomic potential cutoff) were fixed in the z-direction. The outermost layers of the cylindrical surface with a thickness of 1.2 nm were fixed in x and y directions. Periodic boundary conditions were applied in the z-direction and a vacuum layer with a thickness of 4.8 nm was imposed between the periodic images.

Table S1 Potential properties of Mg calculated using the MEAM potential.

| Properties | Experiment/*ab-initio* | MEAM |
|---|---|---|
| $a_0$ (Å) | 3.209[3] | 3.209 |
| $c_0$ (Å) | 5.211[3] | 5.197 |
| $C_{11}$ (GPa) | 63.5[4] | 62.9 |
| $C_{12}$ (GPa) | 25.9[4] | 26.0 |
| $C_{13}$ (GPa) | 21.7[4] | 21.2 |
| $C_{33}$ (GPa) | 66.5[4] | 69.7 |
| $C_{44}$ (GPa) | 18.4[4] | 17.1 |
| $C_{66}$ (GPa) | 18.8[4] | 18.4 |
| $E_{SF\_I1}$ (mJ/m$^2$) | 8.1[4] | 15.1 |
| $E_{SF\_I2}$ (mJ/m$^2$) | 21.8[5] | 30.0 |
| $E_{TB\_\{10\text{-}11\}}$ (mJ/m$^2$) | 85.5[5] | 85.6 |
| $E_{TB\_\{10\text{-}12\}}$ (mJ/m$^2$) | 118.1[5] | 144.0 |
| $E_{GB\_\Sigma 7}$ (mJ/m$^2$) | 298[6] | 335.2 |

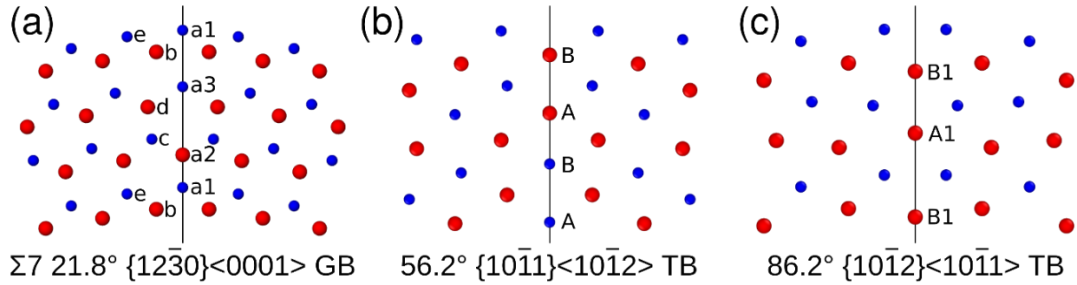

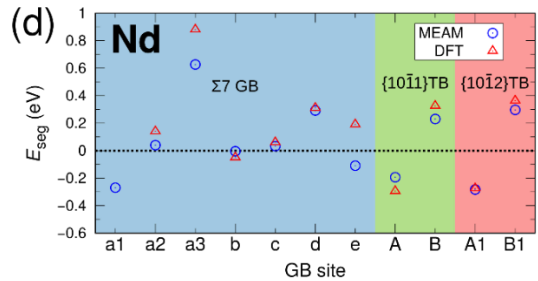

**Fig. S3** Atomic structures of (a) T-type Σ7 21.8° {12$\bar{3}$0}<0001> grain boundary, (b) 56.2° {10$\bar{1}$1}<10$\bar{1}$2> twin boundary and (c) 86.2° {10$\bar{1}$2}<10$\bar{1}$1> twin boundary. Atoms in different atomic layers are distinguished by color and size. (d) Per-site segregation energies at the boundaries in (a-c). Density functional theory (DFT) values are replotted using the data from [6] and [7].

Table S2 Orientation and misorientation data of the simulated grain boundaries.

| Mis. angle (°) | Mis. axis | Orientation (Grain1) Euler Angle | Orientation (Grain2) Euler Angle | GB plane (in Grain1) | GB plane (in Grain2) |
|---|---|---|---|---|---|
| 11.3 | [10$\bar{1}$0] | (174.4 133.7 1.6) | (185.9 141.9 69.6) | ($\bar{1}\bar{1}$47$\bar{4}$) | ($\bar{8}$113$\bar{6}$) |
| 59.3 | [$\bar{3}$121] | (3.5 90.0 327.7) | (46.4 104.3 144.8) | (70$\bar{7}$10) | ($\bar{1}\bar{1}$26) |
| 79.3 | [$\bar{1}\bar{2}$30] | (123.3 90.8 331.9) | (196.4 141.7 243.6) | ($\bar{7}$07$\bar{1}$) | (9$\bar{3}$6$\bar{1}\bar{1}$) |
| 86.5 | [12$\bar{3}$1] | (46.4 104.3 144.8) | (199.0 127.0 310.4) | ($\bar{2}$021) | ($\bar{1}$011) |

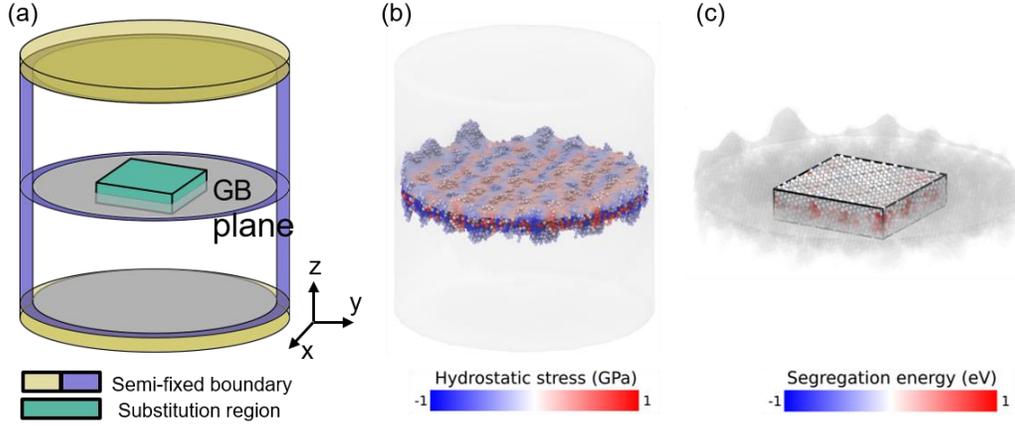

**Fig. S4** (a) Schematic illustration of the cylindrical setup for the atomistic simulations. (b) Atomistic configuration of a 11.3° GB in Mg. Only atoms close to the grain boundary with hydrostatic stress $|\sigma_H|>$ 0.2 GPa are shown. The half-transparent isosurface is constructed using the filtered atoms ($|\sigma_H|>$ 0.2 GPa) with color-coding transferred from the atomic hydrostatic stress. The surface mesh of the simulation sample is half-transparent. (c) Zoomed-in view of the atoms in the substitution region (80 Å × 80 Å × 20 Å) in the center of the simulation sample colored by the segregation energies of corresponding GB sites. Other filtered atoms ($|\sigma_H|>$ 0.2 GPa) and the isosurface are half-transparent.

To characterize the structural rearrangement after each Nd substitution, the per-site mean squared displacement (MSD) was calculated by:

$$\text{MSD} = \frac{1}{N} \sum_{i=1}^{N} \left| x^{(i)}(GB) - x^{(i)}(GB(X)) \right|^2$$

where $x^{(i)}(GB)$ is the coordinate of $i$th atom in a Mg system with a GB, $x^{(i)}(GB(X))$ is the coordinate of $i$th atom in the Mg system with a Nd solute at the GB.

The predicted segregation energy was calculated using the linear elastic model [6]:

$$E_{\text{seg}}^{\text{Elast}} = P_X \Delta V^i$$

$$P_X = B \frac{V_{\text{bulk}}^X - V_{\text{bulk}}}{V_{\text{bulk}}/N}$$

where $\Delta V^i$ is the difference between Voronoi volume of the $i$th GB site and the bulk ($V_{\text{bulk}}/N$), $B$ is the host bulk modulus, $V_{\text{bulk}}^X$ is the volume of the bulk with the solute.

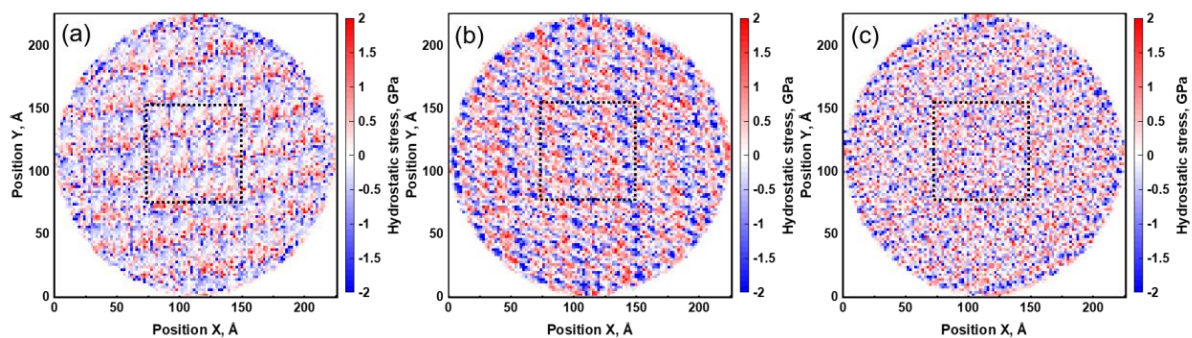

**Fig. S5** Scatter plots of the atomic hydrostatic stress at the (a) 11.3°, (b) 59.3° and (c) 86.5° GBs. The data is divided into 2 Å × 2 Å bins. Only GB sites within a distance of 10 Å from the center of the GB are calculated. The substitution regions are outlined by dashed rectangles.

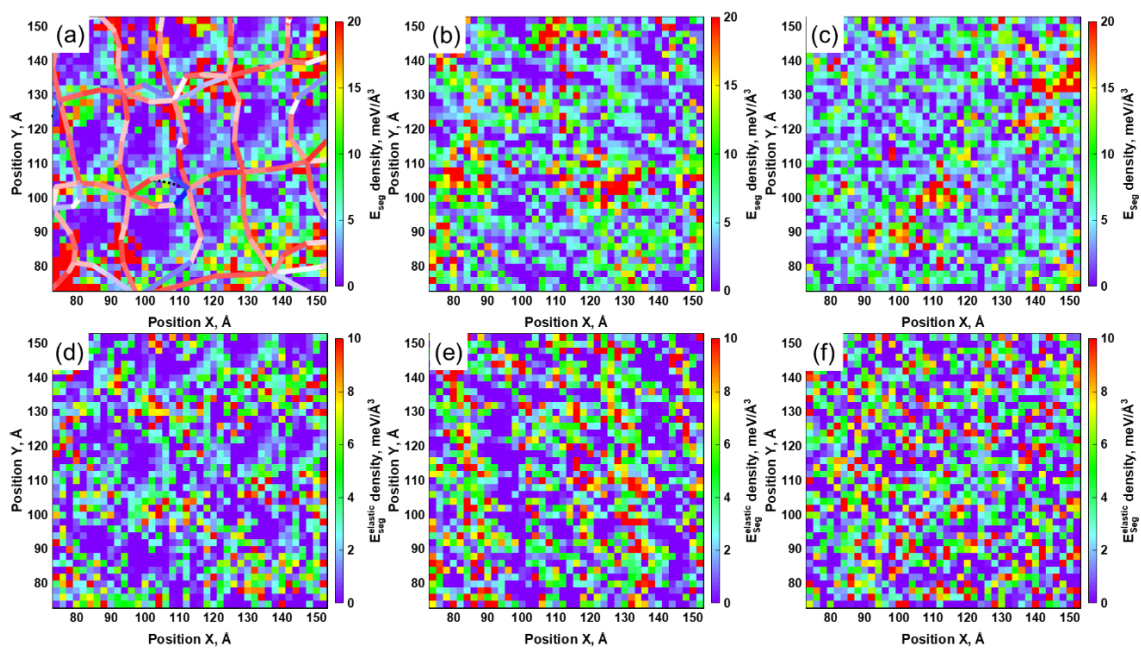

**Fig. S6** Scatter plots of segregation energy density and predicted segregation energy density from the linear elastic model of different GBs: (a) and (d) 11.3°; (b) and (e) 59.3°; (c) and (f) 86.5°. The dashed lines in (a) represent misfit dislocation networks at 11.3° GB. The data is divided into 2 Å × 2 Å bins. Only grain boundary sites within a distance of 10 Å from the center of the grain boundary are calculated.

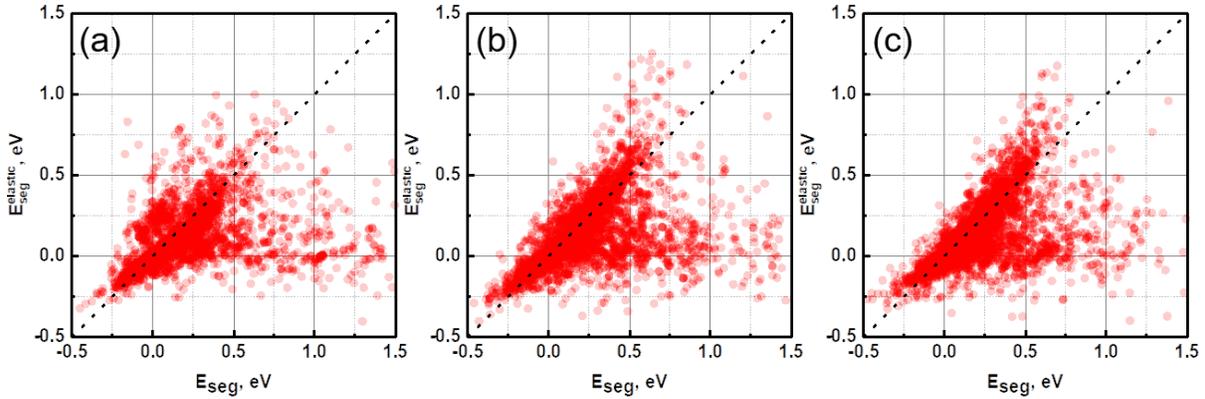

**Fig. S7** Correlation between segregation energies and the predicted segregation energies from the linear elastic model of the (a) 11.3°, (b) 59.3° and (c) 86.5° grain boundaries.

## References

[1]  A.P. Thompson, H.M. Aktulga, R. Berger, D.S. Bolintineanu, W.M. Brown, P.S. Crozier, P.J. in 't Veld, A. Kohlmeyer, S.G. Moore, T.D. Nguyen, R. Shan, M.J. Stevens, J. Tranchida, C. Trott, S.J. Plimpton, LAMMPS - a flexible simulation tool for particle-based materials modeling at the atomic, meso, and continuum scales, Comput. Phys. Commun. 271 (2022) 108171.

[2]  K.-H. Kim, B.-J. Lee, Modified embedded-atom method interatomic potentials for Mg-Nd and Mg-Pb binary systems, Calphad 57 (2017) 55-61.

[3]  C. Barrett, T. Massalski, Structure of Metals, Crystallographic Methods, Principles and Data, (1986).

[4]  G. Simmons, H. Wang, Single crystal elastic constants and calculated aggregate properties, Single Crystal Elastic Constants & Calculated Aggregate Properties 34 (1971).

[5]  Y. Wang, L. Chen, Z. Liu, S.N. Mathaudhu, First-principles calculations of twin-boundary and stacking-fault energies in magnesium, Scripta Mater. 62(9) (2010) 646-649.

[6]  L. Huber, J. Rottler, M. Militzer, Atomistic simulations of the interaction of alloying elements with grain boundaries in Mg, Acta Mater. 80 (2014) 194-204.

[7]  Z. Pei, R. Li, J.F. Nie, J.R. Morris, First-principles study of the solute segregation in twin boundaries in Mg and possible descriptors for mechanical properties, Mater. Des. 165 (2019).

[8]  P. Hirel, Atomsk: A tool for manipulating and converting atomic data files, Comput. Phys. Commun. 197 (2015) 212-219.

[9]  L. Huber, B. Grabowski, M. Militzer, J. Neugebauer, J. Rottler, Ab initio modelling of solute segregation energies to a general grain boundary, Acta Mater. 132 (2017) 138-148.